# Contact holes in vertical electrode structures analyzed by voltage contrast-SEM and conducting AFM

Minseon Gu[1], Moon Seop Hyun,[3] Moonsup Han[1], Gyungtae Kim[3,*], Young Jun Chang[1,2,4,*]

[1]*Department of Physics, University of Seoul, Seoul 02504, Republic of Korea,*

[2]*Department of Smart Cities, University of Seoul, Seoul 02504, Republic of Korea,*

[3]*National NanoFab Center (NNFC), 34141, Republic of Korea,*

[4]*Department of Intelligent Semiconductor Engineering, University of Seoul, Seoul 02504, Republic of Korea*

**Abstract**

Soaring demands of multi-stacked memory devices request urgent development of backside contact electrode technologies, such as high aspect ratio etching, metallization, and inspection methods. Especially the complex metal contact process should be monitored for each manufacturing step to filter the defective samples and to maintain the high yield of production. Among the inspection methods for detecting the electrical connections, there is voltage contrast (VC)-SEM and conducting AFM (C-AFM). In this report, we investigated the two inspection methods for testing designed samples with different contact hole states. The VC-SEM data shows the contrast variation at the contact holes, from which one may discern the contact status with an optimum voltage. The C-AFM results clearly demonstrate a finite electrical current in the connected contact, while a negligible current in the disconnected one. Finally, we discuss insights of using the two methods for analyzing the contact hole technologies with high aspect ratios.

* Corresponding author's e-mails:

Gyungtae Kim    gtkim@nnfc.re.kr,

Young Jun Chang    yjchang@uos.ac.kr

**Introduction**

Our daily lives are heavily intertwined with electronic devices that rely extensively on advanced semiconductor technologies, including smartphones, high-speed networks, cloud computing, and artificial intelligence computing [1]. These technologies have been made possible by the continuous scaling of key device components [2], with critical dimensions reaching a few nanometers in recent years. Thanks to extreme ultraviolet (EUV) lithography and advanced fabrication technologies [3,4], the lateral dimensions of individual devices now approach the nanoscale. In the pursuit of high-density memories, such as dynamic random access memory (DRAM), vertical NAND (V-NAND) flash memory, and resistive switching memory, multi-stacking technologies have become increasingly important [2,5–10]. These multi-stacking approaches necessitate electrical contacts with a high aspect ratio, particularly for backside electrical patterns. Advancements in etching and deposition methods are required to fabricate 3D structures with high aspect ratios, such as 30 for 32-story-tall structures [11]. During the fabrication processes, it is crucial to validate the electrical connection states, distinguishing between the connected condition (open, 'OP') and the disconnected condition (not open, 'NO'), by developing appropriate inspection methodologies. Both speed and precision are key factors for these inspections, as they contribute to enhancing productivity in wafer-scale manufacturing processes.

Various advanced inspection techniques have been explored to test 3D contact holes with high aspect ratios. Electron-based methods, including transmission electron microscopy (TEM) [11], critical dimension-scanning electron microscopy (CD-SEM) [12,13], and voltage contrast-SEM (VC-SEM) [14–16], have been widely developed due to their small beam size and chemical sensitivity, although they require a vacuum environment. Photon-based methodologies, such as ellipsometry [17] and small-angle X-ray scattering (SAXS) [18,19], offer increased sensitivity to diffraction in regularly patterned nanocurie structures. Scanning probe methods, such as atomic force microscopy (AFM) and conducting AFM (C-AFM) [20–23], have emerged as tools for monitoring both topographic and electrical characteristics. These physical probes not only enable topographic inspections but also facilitate active manipulation of nanostructures at scales as small as a few nanometers. However, it is crucial to compare the characteristics of different analytical techniques using similar practical sample structures in order to advance measurement and analysis techniques. Such comparisons can pave the way for developing complementary approaches in the field of nanoscale pattern technology for the next generation.

Here, we conducted an investigation of two examination methods, namely VC-SEM and C-AFM, to assess the electrical connectivity of contact holes in specially prepared test samples. The VC-SEM measurements revealed notable voltage-dependent electron emissions, allowing for effective differentiation between the OP and NO geometries. Fine-tuning of the beam voltage was necessary to achieve optimal distinction for specific materials. On the other hand, the C-AFM tests clearly distinguished current maps between the OP and NO geometries, even when their topographies appeared identical. Finally, we provide insights into the analysis of contact hole states in realistic test samples based on our findings.

**Experimental**

We deposited thick $SiO_2$ on a Si substrate using plasma-enhanced chemical vapor deposition (PECVD) to model the realistic device structure with deep holes. On the $SiO_2$ layer, a hard mask of amorphous carbon layer (1500 nm)/SiON layer (100 nm) was deposited. Then, contact holes were patterned using photolithography with a KrF scanner (S203B, Nikon). The etching of contact holes and masks was conducted using a dry etcher (SCCM, Tokyo Electron Ltd.). The etching process was intentionally stopped with residual $SiO_2$ layers left at the bottom of contact holes to make the electrically floated sample (NO). The thickness of the residual $SiO_2$ was about 100 nm. After the etching of the hole, TiN (10 nm)/W (500 nm) was deposited. TiN was deposited by sputtering (ENDURA-5500, AMAT), and W was deposited by chemical vapor deposition. TiN was selected as a diffusion barrier layer between the Si and W layer along with its superior adhesion property and low contact resistance. Then, chemical mechanical polishing (CMP) was performed to flatten the surface and remove the excess W layer. VC-SEM images were obtained with a Hitachi SU-8230 field-emission SEM. We varied the accelerating voltage from 1 kV to 6.5 kV. The beam current was set to 10 μA. Operating distance was 8–9 mm. Topography and current mapping images were obtained by using AFM (XE-100, Park systems). We used a ElectriMulti75-G rotated-shape tip from BudgetSensors with a radius of 25 nm and a height of 17 μm. We estimated the contact resistance below 1 Ω. We used a preamplifier embedded in the AFM system. A voltage of 2 V was applied to the probe tip. Images were taken with a resolution of 256 × 256 pixels and the scan speed of 1 Hz.

**Results and Discussion**

Fig. 1 describes the schematic outline of the VC-SEM method for the test samples with deep contact hole structures. Two kinds of contact hole geometries in the test samples were prepared with connected (OP) and floated (NO) electrical contacts between the W plugs and the Si substrates. Such different electrical connections make the test samples suitable for the investigations of VC-SEM. For the VC-SEM measurements, a scattered amount of electron beam flux was monitored in terms of the acceleration voltage of the electron beam. In the following, we investigated a detailed comparison of the two different electrical contact states in the test samples.

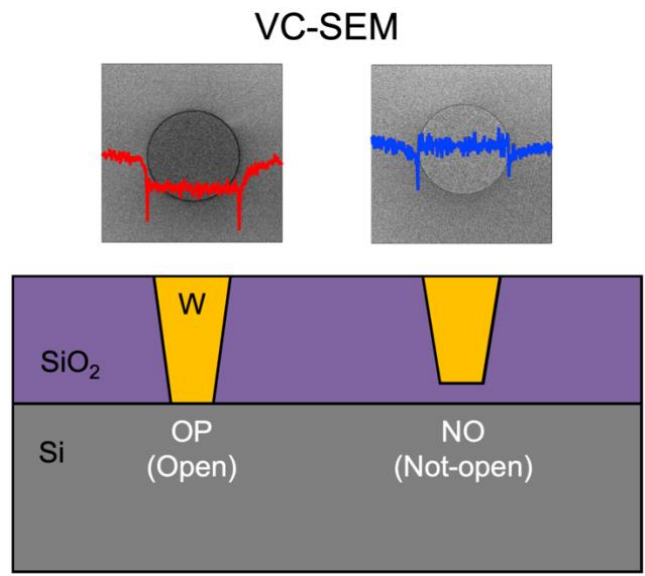

Fig. 1. VC-SEM measurements for the variation of the secondary electron as a function of acceleration voltages. At a certain voltage, the NO electrode shows the same contrast to the side dielectric region (blue profile), while the OP electrode shows depressed contrast (red profile).

Initially, we employed the VC-SEM technique to examine contact hole failure. VC-SEM measures the intensity of the secondary electrons while adjusting the accelerating voltage of the primary electron beam at a constant current (Fig. 1). During this process, there is a partial compensation between the incoming primary electrons and the outgoing secondary electrons, leading to both positive and negative charging on the surface of the sample. The acceleration voltage of the electron beam has the greatest impact on the charging type, which is also influenced by factors such as the primary beam current, sample material, surface topography, and other variables [14]. Although the surface potential can be modulated by applying a bias voltage to the samples, we focus on the grounded sample configuration in this study. Subsequently, the accumulated charges in the dielectric region and the floated W region have distinct effects on the secondary electron signal compared to the connected W region. Leveraging these distinctions, it is possible to identify the optimal range of accelerating voltages that yield the most effective contrast between the OP and NO samples.

Fig. 2(a) and (b) illustrate two sets of VC-SEM images showcasing different acceleration voltages for both the OP and NO samples, respectively. Each SEM image exhibits four contact holes filled with W electrodes, surrounded by $SiO_2$ dielectric areas. As the voltages increase, gradual darkening and brightening of the electrode areas can be observed. However, the contrast variation is more pronounced in the OP sample, particularly in the 3.5–5.5 kV range, compared to the NO sample. Additionally, the contrast variation is less prominent in the side dielectric regions for both samples. This can be attributed to the accumulation of positive charges on the surfaces of the dielectric regions and the floated W electrode regions, which suppresses the emission of negative electrons.

For a qualitative analysis, we extracted the secondary electron signal as a function of the applied voltage. Fig. 2(c) displays the profile plots of secondary electron intensity for the OP sample (red) and the NO sample (blue) along the dashed lines traversing their contact electrodes. As the acceleration voltage increases from 1.0 kV to 6.5 kV, the secondary electron intensity at the electrode area of the OP sample initially decreases, reaches a minimum intensity at 4.5 kV, and then increases. Conversely, the electrode of the NO sample exhibits higher intensities across the entire voltage range compared to the OP sample. It is worth noting that there is minimal intensity variation in the side $SiO_2$ area for both samples. To enhance the discrimination efficiency between the OP and NO samples, the condition of 3.5 kV can be chosen, where the secondary electron intensity is nearly identical to the surroundings for the NO sample. Subsequently, by simply examining whether the electrode area displays suppressed intensity, one can determine the electrical connectivity. Thus, VC-SEM proves to be a valuable tool for identifying the electrical connection states of contact hole electrodes, given the tuned voltage dependence of relative contrast for the specific materials involved.

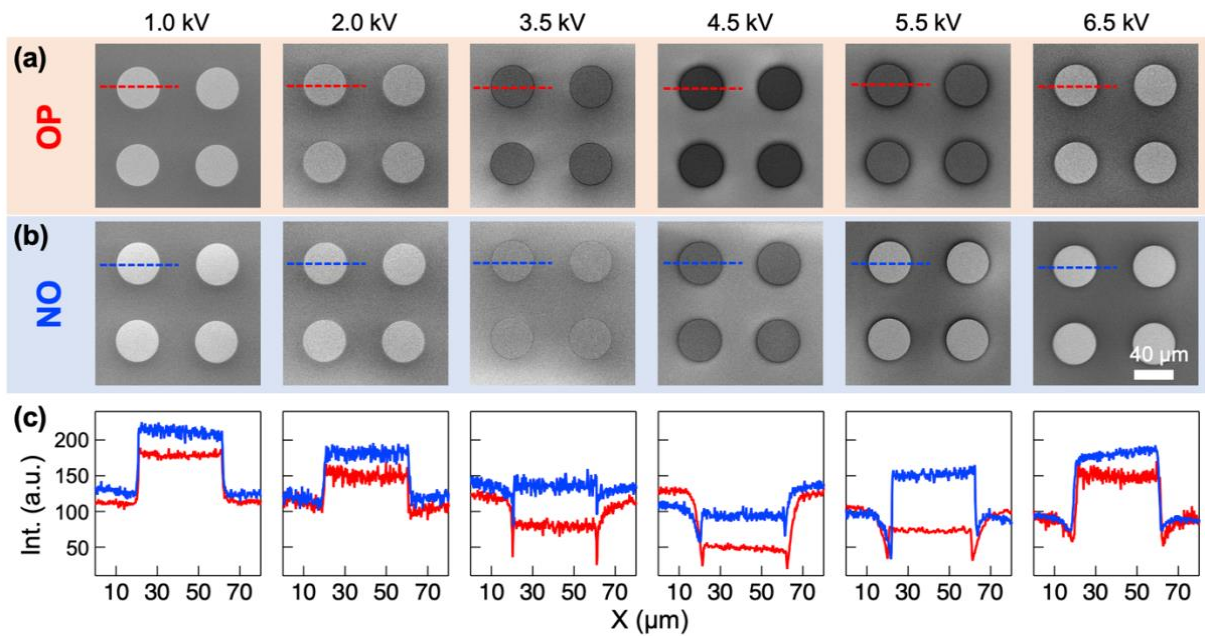

Fig. 2. VC-SEM investigation of the contact hole structures. (a), (b) SEM images taken at varied acceleration voltages (1–6.5 kV) for the OP and NO contact hole samples. (c) Intensity profiles across the contact holes for the corresponding OP (red) and NO (blue) samples show different voltage-dependent behavior. At 3.5 kV, the intensity difference between the contact

electrode and the surrounding is minimal for the NO sample, while that of the OP sample is quite significant. Note that the surrounding areas show only a small contrast variation

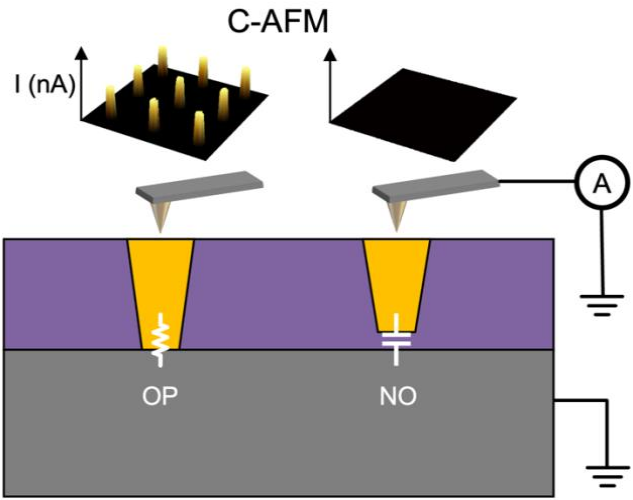

Fig. 3. C-AFM measurements of local current densities, which are sharply varied depending on the contact state of electrodes. Inset images indicate current maps for several contact hole geometries.

Detection of contact hole failure can also be achieved using C-AFM by measuring the electrical current at the surface, as depicted in Fig. 3. In the OP sample, if a W plug is connected to the bottom Si substrate, an electrical current will flow from the Si substrate to the metal-coated C-AFM tips. Conversely, for the floated contact state, the electrical current should be negligible, making it indistinguishable from the dielectric side region. Fig. 4 presents both height and current maps for contact holes of different sizes (1.0 and 0.5 $\mu$m). As the surfaces were flattened using the CMP procedure, the W-filled area appears to be sunken by approximately 40 nm due to the varying polishing rates of W and $SiO_2$. In other words, the topographic images, as indicated by the black curves in the insets, do not differentiate between the OP and NO samples based on the height profiles, regardless of the hole sizes.

Fig. 4 illustrates the current maps (shown in orange), which exhibit a noticeable contrast between the two contact states. In the OP samples, the electrode areas of both contact sizes display similar current values, indicating uniformity. This consistency is further supported by the current profile curves (depicted in red) in the insets. The presence of a non-zero current signal precisely aligns with the locations of the topographic dips, providing direct evidence of both surface structure and electrical conductivity. Conversely, the NO samples exhibit negligible current values across the entire sample area. The disconnected electrode fails to transmit any electrical current, which is consistent with the surrounding dielectric region. Therefore, this comparison serves as unambiguous evidence of how C-AFM examinations effectively differentiate between distinct contact hole states in these model geometries.

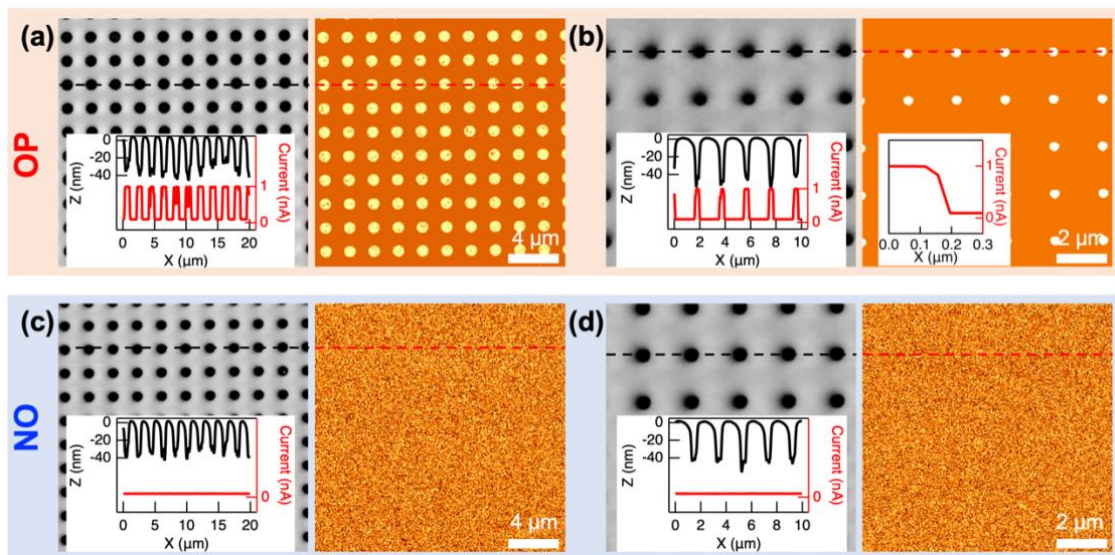

Fig. 4. C-AFM inspection of the contact hole structures. (a-d) Topographic (grey) and current (orange) maps for the OP and NO samples. Topographic maps show very similar dip structures for both OP and NO samples, while the current maps show either peak structures or flat

structures depending on the contact hole states. The insets of the topographic maps show line profiles of height (black) and current (red) values along the dashed lines in the corresponding maps. The inset of (b) displays sharp jump in the expanded current profile near the perimeter of contact hole. The sizes of the holes were 1.0 μm (a, c) and 0.5 μm (b, d).

Our findings provide a direct comparison between two competing methodologies, employing meticulously prepared realistic sample geometries. We further discuss the advantages and limitations of these methodologies for their application in contact hole monitoring procedures. At first glance, VC-SEM appears to be a faster method compared to C-AFM, as the electron beam scanning is quicker than physically probing with tips. However, it is important to note that VC-SEM requires delicate tuning to optimize the secondary electron contrast based on the specific materials and geometries involved. Furthermore, it is crucial to handle surface charging effects effectively in the high vacuum environment, especially when the test samples are not coated with additional metallic layers. In the case of small contact holes, the dielectric region surrounding them can induce significant surface charging phenomena. This is also why we demonstrated VC-SEM monitoring using relatively large contact hole diameters, such as 40 $\mu$m. Considering all these factors, it is important to note that VC-SEM may necessitate a longer probing duration compared to regular SEM scans. This highlights the need for other complementary methodologies that offer ambient compatibility and nanometer resolution.

On the other hand, the results from C-AFM demonstrate a clear capability for classification when applied to the model samples. The bi-functional monitoring of both height and current values proves advantageous for inspection applications. In the inset of Fig. 4(b), the expanded current profile reveals a sharp current jump within the nanometer-scale range at the perimeter of the contact hole. It is noteworthy that the C-AFM, with its ambient application and nanometer-scale lateral resolution, emerges as a promising inspection method alongside SEM-based tools. Further advancements, such as full automation of scanning & tip exchange or high-speed scanning, are currently being developed to overcome the existing limitations of C-AFM and enhance its applicability in semiconductor manufacturing [24,25].

Finally, our thorough comparison of two analytical techniques using similar sample structures offers valuable insights for the development of complementary methodologies. Both VC-SEM and C-AFM demonstrate their usefulness in examining metal-filled contact hole geometries. VC-SEM allows for high-speed scanning, but it requires a high-vacuum environment and careful voltage tuning for optimal results. On the other hand, C-AFM provides high-resolution imaging at ambient pressure but may require automation and a fast-scanning method for improved efficiency. By understanding the advantages and considerations of each technique, we can explore their combined potential for comprehensive contact hole analysis. Let us discuss the future directions of other inspection methodologies. There is currently no appropriate method to assess the condition of contact holes without metallization. The deep holes with high aspect ratios are currently best suited for CD-SEM, but this technique does not provide conclusive evidence of contact connectivity [12,13]. Cross-sectional TEM is a powerful tool, but it is challenging to prepare sliced samples with such deep hole structures, and the process is destructive [26]. X-ray-based spectroscopies can be considered to monitor the etched holes before the deposition of the metal layer, allowing for the detection of the chemical states of the deeply etched holes [27]. Furthermore, the recent advancements in optical spectroscopies and machine learning procedures hold great potential for unprecedented progress in in-line or in-situ inspection methodologies [28–31].

## Conclusion

We conducted a direct comparison between two inspection methods to monitor the electrical connectivity of contact holes, using intentionally designed test specimens that mimic realistic samples. The results from VC-SEM demonstrate the ability to distinguish contact states based on contrast differences, achieved by selecting the optimal acceleration voltage for the specific material and geometry. On the other hand, the C-AFM data exhibit a simpler differentiation of current signals between the samples. Both methods are important tools for monitoring metal-filled contact holes, and our results provide practical examples to guide the development of next-generation semiconductor inspection methodologies. However, there is an urgent need for further innovative methodologies to analyze contact holes without metal filling, in order to meet the growing demands of high-density and high-speed semiconductor devices.

## Declaration of competing interest

The authors declare that they have no known competing financial interests or personal relationships that could have influenced the work reported in this paper.

## Acknowledgements

This research has been supported by the Ministry of Trade, Industry and Energy, Korea and the KSRC (20020927).